\documentclass{article}
\usepackage[a4paper, total={7.2in, 9.5in}]{geometry}
\usepackage[utf8]{inputenc}
\usepackage{blindtext}
\usepackage{ragged2e}
\usepackage{multicol}
\usepackage[font=sf]{caption}
\usepackage{fancyhdr}
\usepackage[pdftex]{hyperref}
\usepackage[dvipsnames]{xcolor}
\usepackage[most]{tcolorbox}
\usepackage{tabularx}
\definecolor{mypurple}{RGB}{70, 10, 128}
\definecolor{mypurple2}{RGB}{217, 179, 255}

\title{AudioUniverse-TourSolarSystem-CAPj}
\author{Chris Harrison}
\date{November 2021}

\begin{document}
\sffamily

 \noindent{\textbf {\LARGE Audio Universe Tour of the Solar System: using sound to make the Universe more accessible}}

\vspace{0.2cm}\noindent Chris Harrison$^{\star}$ (Newcastle University), James Trayford$^{\dagger}$ (University of Portsmouth), Leigh Harrison (Independent), Nic Bonne (University of Portsmouth)\\
\noindent $^{\star}$ christopher.harrison@newcastle.ac.uk, $^{\dagger}$ james.trayford@port.ac.uk

\vspace{0.6cm}\noindent We have created a show about the Solar System, freely available for both planetariums and home viewing, where objects in space are represented with sound as well as with visuals. For example, the audience listens to the stars appear above the European Southern Observatory’s Very Large Telescope and they hear the planets orbit around their heads. The aim of this show is that it can be enjoyed and understood, irrespective of level of vision. Here we describe how we used our new computer code, STRAUSS, to convert data into sound for the show. We also discuss the lessons learnt during the design of the show, including how it was imperative to obtain a range of diverse perspectives from scientists, a composer and representatives of the blind and vision impaired community.

\begin{multicols*}{3}
\noindent\textbf{Introduction}\newline

\noindent Astronomy is successfully communicated through beautiful imagery from artistic impressions and by transforming real data. However, we require powerful telescopes to see most of the light-emitting objects in the Universe and most of the light is not visible to the human eye. Using senses in addition to sight, such as listening, can enable us to discover and appreciate more accurately features in astronomical data (e.g. {\em Díaz-Merced, 2013}). Furthermore, relying exclusively on visual representations is largely inaccessible to people who are blind or vision impaired (BVI).\\

\noindent Recently, there has been increased awareness of these issues. Tactile models provide alternative, BVI-accessible representations of astronomical objects through projects such as Tactile Universe ({\em Bonne et al., 2018})$^{1}$, A Touch of the Universe$^{2}$ and Dedoscopio ({\em Paredes-Sabando and Fuentes-Muñoz, 2021}). There has also been increased interest in using sound for both astronomy research and communication (e.g. {\em Cooke et al., 2019}; {\em Bieryla et al., 2020}; {\em Elias Elmquist et al., 2021}). Objects/phenomena can be represented through sound via either an ‘artistic’ approach or the conversion of data (‘sonification’).\\  

\noindent We have created a new project called ‘Audio Universe’ which will provide a collection of tools and resources using sound for astronomy research and communication. To launch this project we present our new show ‘Audio Universe: Tour of the Solar System’. In this article we describe our design approach, the lessons learnt and some technical details of the sound design and sonification. The show can be accessed through our website from 7th December 2021$^{3}$.\\

\begin{figure*}
\centerline{\includegraphics[width=0.9\textwidth]{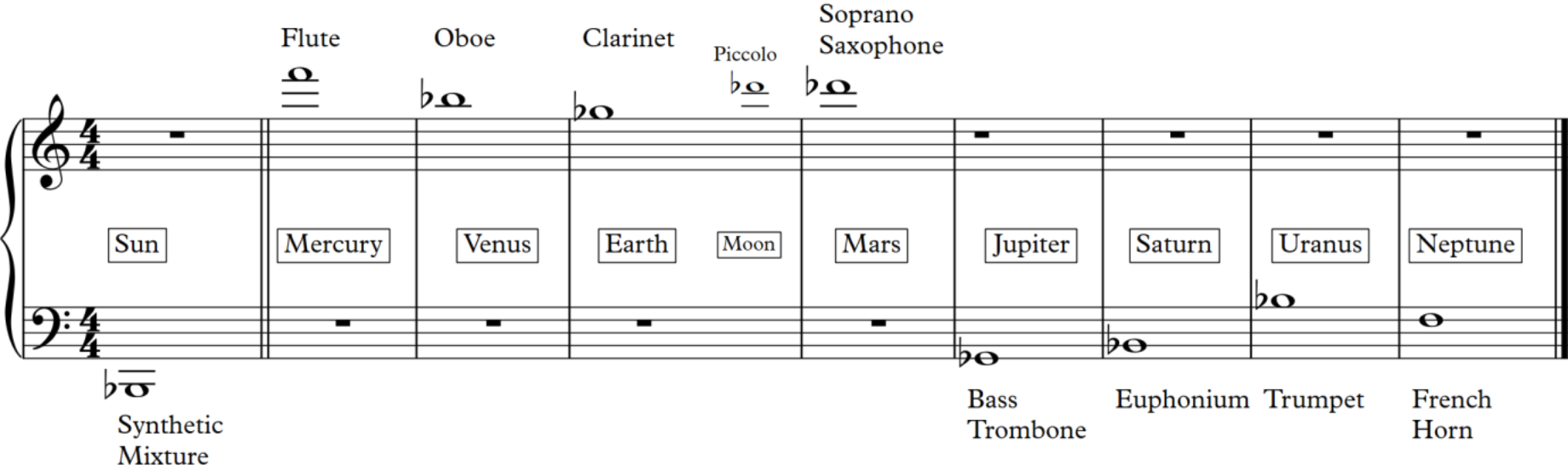}}
\caption{Musical score to show the pitches that were used to represent each of the Sun, the eight planets and the Moon. The information is also presented in tabular form in Table 1. }
\end{figure*}

\noindent\textbf{Pilot Show and Learning Experience}\newline

\noindent Our new show was preceded by a pilot BVI-accessible show called ‘Dark Tour of the Universe’$^{4}$. This 1 hour show explained astronomical objects and phenomena using a combination of narration, sonifications and tactile models. Visuals were intentionally excluded. We considered this an overall success with positive feedback, including from qualified teachers of visually impaired children (QTVIs). One of which commented: {\em “It was so good to be the focus of a presentation, rather than an ‘add on’ with adapted resources or being part of something that isn't wholly accessible”} (Rachel Lambert, Newcastle Children's Vision Team). However, three lessons that we learnt were:\\

\noindent i. {\em It is crucial to provide clear context to the audience before playing the audio.} 
Audiences are not used to the idea of representing light or data with sound and it is difficult for them to quickly interpret the sonifications. Therefore, we decided we should include sonification ‘training’ for the audience to introduce the concept of turning light into sound. Additionally, we should ensure that the narration gives sufficient information for the audience to interpret what they hear.\\

\noindent ii. {\em Visual representations need not be excluded.}
Combining visuals with audio provides another method of communication for people with varying levels of sight. We received this feedback from our partially and fully sighted audience members. Including visuals, even when the target audience is BVI, is also advocated by blind astronomer Enrique Pérez Montero ({\em Pérez Montero, 2019}).\\

\noindent iii. {\em Using tactile models as a mandatory part of the show limits reach and dissemination.}
We received many requests to share our pilot show from planetariums and other communicators. However, whilst audio files could easily be shared, tactile models are more prohibitive due to the challenges and costs of manufacturing them in large quantities. Therefore, we concluded that future shows should be available online as audio-visual files and tactile models should be considered an optional addition.\\ 

\noindent\textbf{Design and Lessons Along the Way}\\

\noindent Building on our pilot, we set out to create a 35 minute audio-visual show about the Solar System that would be an enjoyable and educational experience from the soundtrack alone. The show was to be created in both full-dome planetarium format (with surround sound) and in flatscreen format (with stereo sound). Here we outline our design approach and our lessons learnt.\\

\noindent{\em Concept and Design Approach}\\

\noindent The audience imagine themselves inside a spacecraft that is equipped with a ‘sonification machine’ that turns light into sound. To ‘train’ the audience ‘pre-flight tests’ are conducted, where lights are played around the spacecraft and the audience can listen to their sonification. The audience are then taken to various locations to learn about the stars, Sun, Moon and planets. At each point they are presented with sounds as well as visuals.\\

\noindent Unlike most shows, we created the soundtrack before the visuals. At various stages we sought feedback from members of the BVI community (including children and adults), BVI astronomers, and QTVIs. This was crucial and led to various changes to our initial designs. For example, we originally used ‘spacey’ synthesizer sounds to represent the planets, but the suggestion from our feedback groups was to represent them with more recognisable and easily distinguishable musical instruments.\\

\noindent We also included an inspiring role model for our BVI audience, by using real blind astronomers as the ‘tour guides’. These are Dr Nic Bonne and Dr Enrique Pérez Montero in the English and Spanish versions, respectively. For translations into other languages one of these is chosen.\\ 

\noindent{\em Balancing Science and Music}\\

\noindent A challenge when designing the show was to find the correct balance between scientific accuracy and producing a musically-pleasing result. Based on our consultations from the feedback groups it became clear that something that is musically pleasing to listen to was very important for the show to be engaging and enjoyable. The science being discussed should still be represented with the sounds, but it was not always necessary to accurately tie this directly to the data/properties of interest (also see {\em Elias Elmquist et al., 2021}). Therefore, a composer was recruited to help with creating the soundtrack.\\ 

\vspace{0.2cm}\noindent {\small Table 1: The musical notes, using International Pitch Notation, and the instruments used for the Sun, the eight planets and the Moon.}
\begin{center}
\begin{tabular}{ccc}
\hline
\textbf{Object} & \textbf{Note} & \textbf{Instrument} \\
\hline
Sun & Bb1 & Synthetic Mix \\
Mercury & F6 & Flute \\
Venus & Bb5 & Oboe \\
Earth & Gb5 & Clarinet \\
Mars & Db6 & Soprano Sax. \\
Jupiter & Gb2 & Bass Trombone \\
Saturn & Bb2 & Euphonium \\
Uranus & Bb3 & Trumpet \\
Neptune & F3 & French Horn \\
Moon & Db6 & Piccolo\\
\hline
\end{tabular}
\end{center}
\vspace{0.5cm}

\begin{figure*}
\centerline{\includegraphics[width=0.95\textwidth]{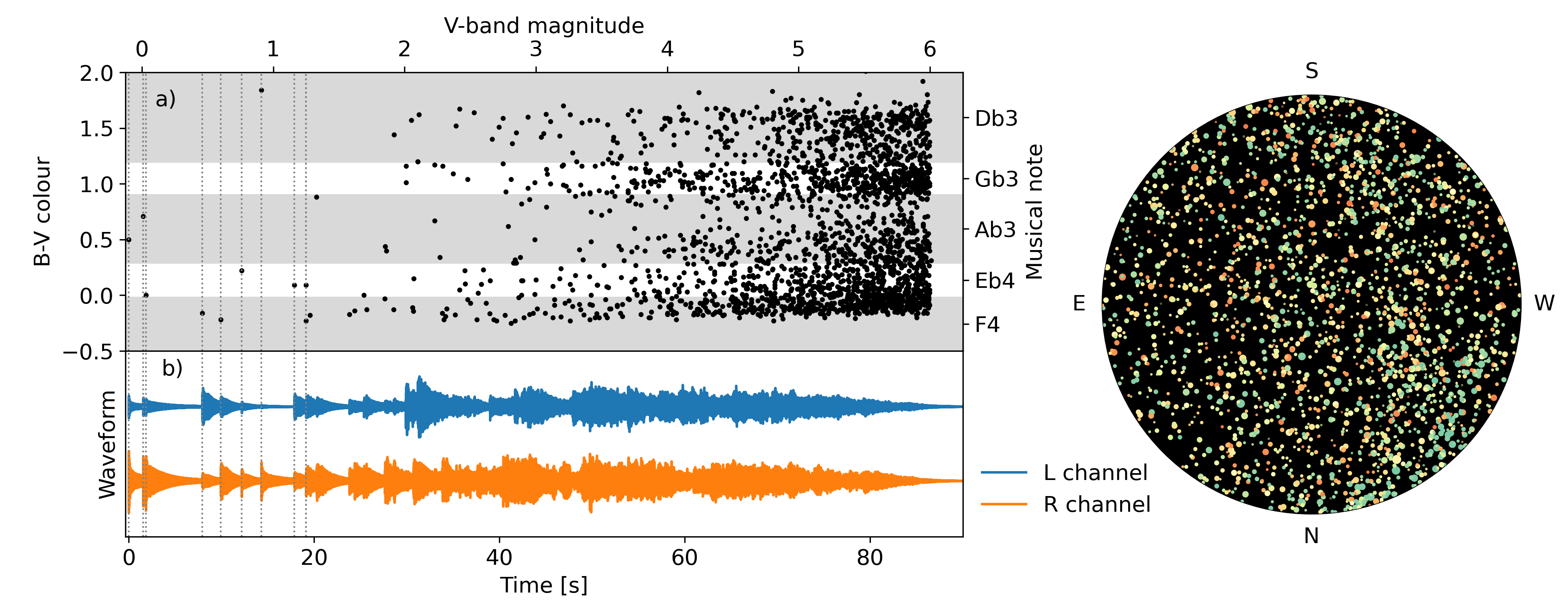}}
\caption{Diagram demonstrating our sonification of the ‘stars appearing’. Panel a) shows the mapping of V-band magnitude (top axis) and B-V colour (left axis) to triggering time in the audio sequence (bottom axis) and musical note (right axis), respectively. The aligned Panel b) shows the waveform produced for a stereo setup, and the triggering times of the 10 brightest stars (dotted lines). The right panel shows the stellar sky chart, with point size and colour indicating brightness and B-V colour, respectively. In our sonification the observer faces south, with the left and right audio channels corresponding to the east and west cardinal directions, respectively. Our audio-visual representation is available here: \url{https://youtu.be/5HS3tRl2Ens}.}
\end{figure*}

\begin{figure*}
\centerline{\includegraphics[width=0.95\textwidth]{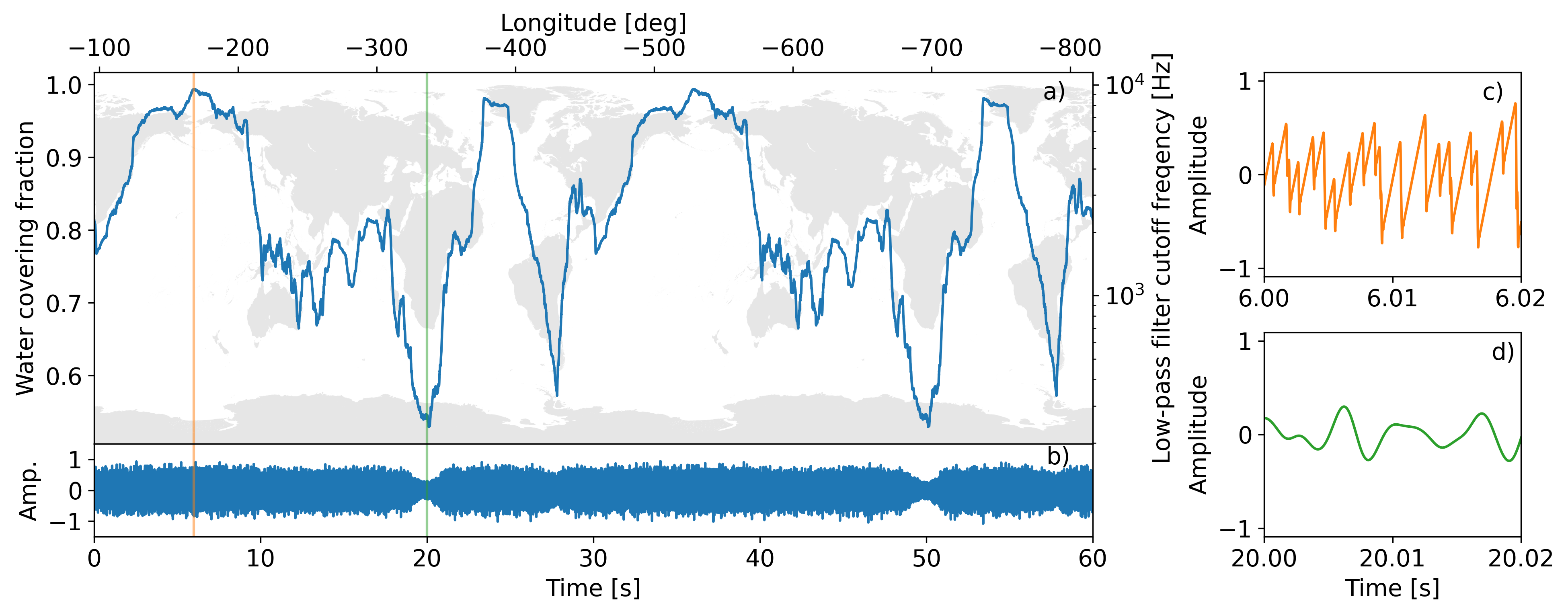}}
\caption{Diagram demonstrating our data sonification of the Earth’s rotation. Panel a) shows water covering fraction (left axis) versus longitude (top axis) over two Earth rotations with a world map projection as a grey underlay. These values are used to calculate the low-pass filter cutoff frequency for the sonification (right axis). Panel b) shows the waveform of the sonification as a function of time. Panels c) and d) demonstrate the effect of the filter on the waveform by zooming into 20 milli-second windows around longitudes where the water covering fraction is approximately highest and lowest, respectively (indicated by vertical lines of corresponding colour in Panel a). Our audio-visual representation is available here: \url{https://youtu.be/-1jVH9-v7yM}.}
\end{figure*}

\noindent The starting point was to find pitches to represent the Sun and eight planets.  We wanted the smaller rocky planets to be represented by higher pitches and the gas giants to be represented as lower pitches. However, the exact choice of pitches was based on musical judgement such that when the Sun and all the planets were heard together, it would produce a pleasing sound.  The Sun acts as a musical “pedal point” for the planets’ pitches which create a blend of a Bb minor chord and a Gb major chord (see Table 1 and Fig. 1 for the exact pitches). Based on the feedback we received, the distinctive timbres of woodwind and brass instruments were chosen to represent the rocky planets and gas planets, respectively. The sound of the Sun is a composite of a pure tone, noise and pre-recorded fire crackle sounds to give the overall impression of a powerful ‘burning’ object. The same musical notes and instruments are used to form the motif of the background music. This consistency creates an overall musical effect for the whole show.\\

\noindent \textbf{The STRAUSS Code and Sonifications}\\

\noindent To create the sonfications we utilised our new code {\em \textbf{S}onification \textbf{T}ools and \textbf{R}esources for \textbf{A}stronomers \textbf{U}sing \textbf{S}ound \textbf{S}ynthesis} (STRAUSS; {\em Trayford et al., in prep.}). Briefly, this python code$^{5}$ allows the user to synthesise sounds or manipulate pre-recorded audio samples with properties (e.g. pitch, volume, timbre) that are mapped from input data.\\ 

\noindent You can also move objects in artificial space around an ‘observer’ and the code will correctly calculate the relative volumes for the different channels of different speaker setups (e.g. stereo or 5.1 systems). For example, we used this to assign a sound to the Moon and the planets and then create the illusion, through sound, that they orbit around the audience. However, based on our feedback groups, we added a volume envelope to emphasise this orbiting effect. Specifically, when an object is in front of the ‘observer’ it sounds louder and when it is behind it sounds quieter. Below, we briefly describe how STRAUSS was used to create two more sonifications for the show.\\

\noindent{\em Stars Appearing}\\

\noindent The audience ‘listen’ to stars that appear around them at the European Southern Observatory’s Very Large Telescope (VLT). The data we used are presented Fig. 2, which are the magnitudes, colours and coordinates of stars as viewed from the VLT on the 13th September 2019. We only considered stars with V-band magnitudes $<$6, to roughly correspond to the detection limit of the human eye.\\ 

\noindent Each star is represented by a single note on a glockenspiel from one of five pitches: Db3, Gb3, Ab3, Eb4 or F4, with the choice of note based on the star’s colour (specifically, the difference in the star’s B and V magnitude)$^{6}$. The reddest stars are assigned the lowest notes and the bluest stars the highest notes. During the sequence each star is heard in an order based on its magnitude, with the brightest stars sounding first and the faintest sounding last. This represents how brighter stars appear first to the human eye after sunset.\\

\noindent The stars’ positions were used to determine in which speaker(s) they should be heard. For example, stars directly in front of the ‘observer’ sound in the front speakers for the surround sound version, or equally in the left and right ear for the stereo version.\\ 
 
\noindent{\em Earth Spinning}\\

\noindent We wanted to sonify sunlight bouncing off the spinning Earth through changing timbre as the Sun passes over water (a “brighter” sound) or land (a “darker” sound). For this, we used data of the covering fraction of water as a function of longitude. The data are from the GEBCO 2021 bathymetry, which assigns water or land to each cell of a 15x15 arcsecond grid across the Earth$^{7}$ (Fig. 3). To create the sonification, we started with a sustained musical chord, using notes Gb3, Db4, E4 and B4. Each note was created from a set of three sawtooth oscillators combined at frequencies on, 2\% above and 2\% below the target pitch. These choices provide a harmonically rich sound, which is then manipulated by filtering out frequencies (i.e. subtractive synthesis) based on the water covering fraction data.\\ 

\noindent The longitude and water covering fraction (Fig. 3a, upper and left axes) were mapped directly to the time in the sequence and the filtering cut-off frequency of the chord, respectively (bottom and right axes). The cut-off frequency, above which frequencies are attenuated, was calculated from the water covering fraction using a logarithmic scale. A low-pass Butterworth filter ({\em Butterworth, 1930}) with a 24dB roll-off was used. The conversion of water fraction to frequency cut-off can be seen by comparing the left and right y-axes of Fig. 3a. We note the more jagged, harmonically rich waveform representing a longitude over the pacific ocean (Fig. 3c) relative to the smoother waveform representing a longitude over Europe and Africa (Fig. 3d). The filtering mainly changes the timbre of the sound but a secondary effect on volume is achieved in that the land-dominated regions sound the quietest (see waveform in Fig. 3b).\\ 

\noindent\textbf{Conclusions}\\

\noindent Our BVI-accessible ‘Audio Universe: Tour of the Solar System’ is released for both planetariums and home viewing from December 7th 2021$^{3}$, initially in English, Spanish and Italian, with other languages to follow. The next steps will be to use individual sections of the show, in combination with tactile models, to create a series of accessible interactive workshops. In this article we have presented many of the lessons learnt from our pilot show and developing this new show, including the importance of consultation with groups with a variety of experiences and expertise. We hope that future astronomy shows consider how their soundtracks can be used effectively to aid accessible communication and that multi-sensory techniques become the norm when communicating astronomy to the general public.

\vspace{0.5cm}\noindent\textbf{Notes}\newline

\small{
\noindent $^{1}$ Tactile Universe project: \url{https://tactileuniverse.org/}. 

\vspace{0.1cm} 
\noindent $^{2}$ A Touch of the Universe project: \url{https://www.uv.es/astrokit/}.

\vspace{0.1cm}  
\noindent $^{3}$  Audio Universe web pages and links to audio-visuals: \url{https://www.audiouniverse.org/}.

\vspace{0.1cm} 
\noindent $^{4}$  Summary of ‘Dark Tour of the Universe’, including example sounds: \url{https://www.eso.org/public/announcements/ann19045/}.

\vspace{0.1cm}
\noindent $^{5}$  STRAUSS Code: \url{https://github.com/james-trayford/strauss}.

\vspace{0.1cm} 
\noindent $^{6}$  We note that the version of the stars appearing sequence that is used in the full show did not use an exact accurate mapping of colour to pitch in the way described here. We have since updated the code for this accuracy mapping.

\vspace{0.1cm} 
\noindent $^{7}$  Data obtained from GEBCO Compilation Group, 2021, “GEBCO 2021 Grid”, (doi:10.5285/c6612cbe-50b3-0cff-e053-6c86abc09f8f).

}

\vspace{0.5cm}\noindent\textbf{References}\\
{\small 

\noindent\textbf{Bieryla, A. et al.}, 2020, ‘LightSound: The Sound of An Eclipse’, {\em CAPjournal}, \textbf{no. 28}, p. 3\vspace{0.1cm}

\noindent \textbf{Bonne, N. , Gupta, J., Krawczyk, C. and Masters, K.}, 2018, ‘Tactile Universe makes outreach feel good’, {\em Astronomy \& Geophysics}, \textbf{59a}, p. 1.30\vspace{0.1cm}

\noindent \textbf{Butterworth, S.}, 1930, ‘On the theory of filter amplifiers’, {\em Wireless Engineer} \textbf{7(6)}, p. 536\vspace{0.1cm}

\noindent \textbf{Cooke, J., Díaz-Merced, W., Foran, G., Hannam, J. and Garcia, B.}, 2019, ‘Exploring Data Sonification to Enable, Enhance, and Accelerate the Analysis of Big, Noisy, and Multi-Dimensional Data’, {\em Southern Horizons in Time-Domain Astronomy Proceedings IAU Symposium} \textbf{No. 339}, p. 251\vspace{0.1cm}

\noindent \textbf{Díaz-Merced, W.}, Sound for the exploration of space physics data, {\em Ph.D. Thesis, Computer Science, Univ. Glasgow} \textbf{258}, 2013\vspace{0.1cm}

\noindent \textbf{Elias Elmquist, E., Ejdbo, M., Bock, A. and Ronnberg, N.}, 2021, ‘Open Space Sonification: Complimenting Visualization of the Solar System with Sound’, {\em Proceedings of the International Community of Auditory Displays 2021}\vspace{0.1cm}

\noindent \textbf{Paredes-Sabando, P. and Fuentes-Muñoz, C.}, 2021, ‘Dedoscopio Project: Making Astronomy Accessible to Blind and Visually Impaired (BVI) Communities Across Chile’, {\em CAPjournal} \textbf{29},  p. 27\vspace{0.1cm}

\noindent \textbf{Pérez Montero, E.}, 2019, ‘Towards a more inclusive outreach’, {\em Nature Astronomy} \textbf{3}, 114\\

}

\vspace{0.1cm}\noindent\textbf{Acknowledgements}\\

\noindent Audio Universe: Tour of the Solar System was part funded by a Science and Technology Facilities Council Spark Award (Grant: ST/V002082/1) and two Royal Astronomical Society Education and Outreach Small Grants. We are very grateful for the valuable contributions from: Steve Toase, Aishwarya Girdhar, Rachel Lambert, Amrit Singh, Anita Zanella, Jeff Cooke, Phia Damsma, Garry Foran, Rubén Garcia-Benito, Miranda Jarvis, Theofanis Matsopoulos, Liz Milburn, Enrique Pérez Montero, Stefania Varano, The Institute of Cosmology and Gravitation at the University of Portsmouth, Newcastle Children’s Vision Team, The VIEWS group Newcastle and The Great North Museum: Hancock.

\end{multicols*}

\end{document}